\documentclass[11pt,a4paper]{article}
 \def\bea{\begin{eqnarray}}
 \def\eea{\end{eqnarray}}
 
 \def\beq{\begin{equation}}
 \def\eeq{\end{equation}}
 \def\ba{\beq\new\begin{array}{c}}
 \def\ea{\end{array}\eeq}
 \def\be{\ba}
 \def\ee{\ea}

 \makeatletter
 \newdimen\normalarrayskip 
 \newdimen\minarrayskip 
 \normalarrayskip\baselineskip \minarrayskip\jot
 \newif\ifold \oldtrue \def\new{\oldfalse}
 \def\arraymode{\ifold\relax\else\displaystyle\fi} 
 \def\eqnumphantom{\phantom{(\theequation)}} 
 \def\@arrayskip{\ifold\baselineskip\z@\lineskip\z@
 \else \baselineskip\minarrayskip\lineskip2\minarrayskip\fi}
 \def\@arrayclassz{\ifcase \@lastchclass \@acolampacol \or
 \@ampacol \or \or \or \@addamp \or \@acolampacol \or
 \@firstampfalse \@acol \fi \edef\@preamble{\@preamble \ifcase
 \@chnum \hfil$\relax\arraymode\@sharp$\hfil \or
 $\relax\arraymode\@sharp$\hfil \or
 \hfil$\relax\arraymode\@sharp$\fi}}
 \def\@array[#1]#2{\setbox\@arstrutbox=\hbox{\vrule
 height\arraystretch \ht\strutbox depth\arraystretch \dp\strutbox
 width\z@}\@mkpream{#2}\edef\@preamble{\halign \noexpand\@halignto
 \bgroup \tabskip\z@ \@arstrut \@preamble \tabskip\z@ \cr}%
 \let\@startpbox\@@startpbox \let\@endpbox\@@endpbox
 \if #1t\vtop \else \if#1b\vbox \else \vcenter \fi\fi \bgroup
 \let\par\relax
 \let\@sharp##\let\protect\relax
 \@arrayskip\@preamble}
 \def\eqnarray{\stepcounter{equation}%
 \let\@currentlabel=\theequation
 \global\@eqnswtrue \global\@eqcnt\z@ \tabskip\@centering
 \let\\=\@eqncr
 $$%
 \halign to \displaywidth\bgroup
 \eqnumphantom\@eqnsel\hskip\@centering
 $\displaystyle \tabskip\z@ {##}$%
 \global\@eqcnt\@ne \hskip 2\arraycolsep
 $\displaystyle\arraymode{##}$\hfil \global\@eqcnt\tw@ \hskip
 2\arraycolsep $\displaystyle\tabskip\z@{##}$\hfil
 \tabskip\@centering &{##}\tabskip\z@\cr}
 \begingroup\ifx\undefined\newsymbol \else\def\input#1 {\endgroup}\fi

\begin{document}
 \setcounter{footnote}{1}
 \def\thefootnote{\fnsymbol{footnote}}
\begin{center}
\hfill ITEP/TH-62/02\\
\hfill hep-th/0212005
\end{center}

 \begin{center}
 \vspace{0.3in} {\Large\bf Anticommutativity Equation in Topological Quantum Mechanics }
 \end{center}
 \centerline{{\large Vyacheslav Lysov}\footnote {
 e-mail: sllys@gate.itep.ru }}
 \centerline{{\normalsize \it \\  } }
\centerline{{\normalsize \it Institute of Theoretical and
Experimental Physics, 117259, Moscow, Russia  } }
\centerline{{\normalsize \it and } } \centerline{{\normalsize \it
Moscow Institute of Physics and Technology, 141700, Moscow,
 Russia}}
\abstract{\footnotesize
We consider  topological quantum mechanics as an example of topological field theory
and show that its special properties lead to numerous interesting  relations
for topological corellators  in this theory. We prove that the generating function $\mathcal{F}$
 for thus corellators satisfies  the anticommutativity equation
$(\mathcal{D}- \mathcal{F})^2=0$. We show that the commutativity equation $[dB,dB]=0$ could be considered
as a special case of the anticommutativity equation.}
\begin{center}
\rule{5cm}{1pt}
\end{center}

 {\bf 1.}
  During the last   two decades there has been much interest to quantum field
theories whose special corellators do not depend on coordinates  and
metric.
 These theories are called topological[1]. The
most celebrated examples include Chern-Simons theory [2], $N=2$
twisted gauge theories[3], topological sigma models[4]. Here we consider
yet another, much simpler example of topological theory which is a
subsector of  supersymmetric quantum mechanics. We call it
topological quantum mechanics.

An explicit construction of many topological theories is given by
a BRST-like symmetry operator $Q$, $Q^2=0$ so that the energy momentum tensor of the theory
has the special form [3]
\be
T_{\mu\nu}=\{Q,G_{\mu\nu}\}
\ee
where $G_{\mu\nu}$ is some tensor.
This formula leads to many interesting corollaries. In particular, it
makes corellators of $Q$-closed operators  independent of metric and
coordinates, that is topologically invariant.

In the case of one dimensional theory, the energy momentum tensor
has only one component and is equal to the Hamiltonian
\be\label{2}
 H= T_{00}  =\{Q, G_{00} \}  =\{Q, G_+ \} \ee
In topological quantum mechanics, the objects of study are
corellators $ <\Phi_{A_1}(t_1)... \Phi_{A_n}(t_n)> $ of $Q$-closed
operators. They would naively depend of $n$ times
$t_{1}<...<t_{n}$. However, since the energy momentum tensor is
anticommutator of $Q$ and $G_+$, the corellator actually does not
depend on $t$'s and is given by a factorization formula \be
<\Phi_{A_1}(t_1)...\Phi_{A_n}(t_n)>=
<0|e^{-t_1H}\Phi_{A_1}e^{(t_1-t_2)H}\Phi_{A_2}...e^{(t_{n-1}-t_n)H}\Phi_{A_n}e^{t_nH}|0>=\\
\lim_{(t_k-t_{k-1}) \rightarrow \infty}
<0|\Phi_{A_1}e^{(t_1-t_2)H}...e^{(t_{n-1}-t_n)H}\Phi_{A_n}|0>=
<\Phi_{A_1}>...<\Phi_{A_n}> \ee
 It means that it is sufficient  to study the
corellators of  one operator $\Phi_A$ in the theory with the
deformed supercharge $Q$
 \be
 Q \rightarrow Q+ \sum\Phi_AT_A =Q+\Phi
\ee
 Later we shall see that this one-point corellator is a total
derivative of $\mathcal{F}$ on the space of coupling constants
$T_A$ \be\label{cor}
\partial_A\mathcal{F}=<\Phi_A>_{\hbox{deformed}}=<T\left\{\Phi_Ae^{\int \{\Phi,G_+\}dt}\right\}>
\ee
where $T\{...\}$ stands for the chronological ordering.
In what follows, we carefully formulate the theory and realize that the function $\mathcal{F}$
satisfies an interesting quadratic differential equation which we call
anticommutativity equation.

\noindent
{\bf 2.}
  Thus, we consider the simplest example of topological field theory, topological
quantum mechanics. As usual, there is a nilpotent
  symmetry operator $Q$ and one can introduce the Hamiltonian in the form $(2)$.
  All  operators act on the space $V$ with the following properties
  \be
  V=V_1 \oplus V_0 \\
  HV_0=0 \    \ , \    \ QV_0=0 \\
  H>0  \  \ \hbox{on}\  \ V_1
  \ee
  This physically means that $V$ is a space of states of our system, $V_0$ is a space of the vacuum states .
  We assume that the kernel $ V_0 $ of the Hamiltonian consists of $Q$-closed states, and
  all non-zero energies have strictly positive real part.
 In this theory we study the corellators of the form (\ref{cor}), which are the vacuum matrix elements, i.e.
the operators from $ Hom (V_0,V_0) $ and can be written in the following
form\footnote{Here we used the properties $G_+^2=0$ and $G_+V_0=0$ that allow one to interpret $F$
as a corellator in topological theory, but they are not necessary for our main result.}
  \be
 F_{\alpha \beta }^{(k)}=\int_0^\infty  ... \int_0^\infty  <0_\alpha|
\Phi G_+ e^{-t_1H} \Phi G_+ e^{-t_2H} ...\Phi |0_\beta>  dt_1..dt_{k-1}=\\
=\frac{1}{k!}<T\left\{\Phi(0) (\int_{-\infty}^{+\infty}\{G_+,\Phi(t)dt\})^{k-1}\right\}>_{\alpha \beta}
\ee

We want to represent this operator from $Hom(V_0,V_0)$ as an
operator from $Hom (V,V)$. The way to do this is to insert
$\Pi_0$'s (projector on to $V_0$ )  at the beginning and at the end.
 After doing this, one finally obtains the object  needed
 \be
  \mathcal{F}^{(k)}=\int_0^\infty  .. \int_0^\infty
  \Pi_0 \Phi G_+ e^{-t_1H} \Phi G_+ e^{-t_2H} ....\Phi \Pi_0 dt_1..dt_{k-1}
  \ee
  The $\Phi(0)\equiv\Phi $ is
  \be
 \Phi(0)= \Phi = \sum\Phi_A T_A \\
\ee
Our theory is Euclidean, which means that the evolution of $\Phi$ in time can be described
by the following formula
\be
\Phi(t)=e^{-tH}\Phi(0)e^{tH}
\ee
$\Pi_0$ being a projector onto the vacuum states. One can
rewrite $\Pi_0 $ in terms of $H$
\be
 \Pi_0=\lim_{t\rightarrow \infty } e^{-tH}
\ee
It commutes with $Q$. Therefore,
\be
Q\Pi_0=\Pi_0Q=0 \\
<0_\alpha |Q=<0_\alpha|\Pi_0Q=0
\ee
Our operators can be both even and odd, therefore, we  introduce a superalgebra
to describe their properties, our coupling constants $T_A$ being graded too. Their algebra
is
\be
  \{T_A,T_B\}_s=0
\ee
Where $\{.,.\}_s$  stands for supercommutator.
In order to get  interesting properties of the
corellators, let us consider a special set of operators that
satisfy the following algebra \be\label{relf}
  \{\Phi_A,\Phi_B\}_s=C_{AB}^K \Phi_K \\
  \{Q,\Phi_A\}_s=0
  \ee

\noindent
{\bf 3.}
  Algebra of operators $\Phi$ {\bf generates numerous commutation relations} for $\mathcal{F}^{(k)}$
\be\label{f1}
\{\mathcal{F}^{(1)},\mathcal{F}^{(1)}\}=T_A T_B
C_{AB}^K\partial_{K}\mathcal{F}^{(1)} \\
2\{\mathcal{F}^{(2)},\mathcal{F}^{(1)}\}=T_A T_B
C_{AB}^K\partial_{K}\mathcal{F}^{(2)} \\
2\{\mathcal{F}^{(1)},\mathcal{F}^{(3)}\}+\{\mathcal{F}^{(2)},\mathcal{F}^{(2)}\}=
T_A T_B C_{AB}^K\partial_{K}\mathcal{F}^{(3)} \\
...
\ee
 which
  can be written in a compact form in terms of the $Hom(V_0,V_0)$ valued  generating function $\mathcal{F}$
   \be
  \mathcal{F}=\sum {\mathcal{F}^{(k)}}
  \ee
as
\be
  \{ \mathcal{F}, \mathcal{F}\}=2\mathcal{D}\mathcal{F}
  \ee
   Here we introduced the derivative $\mathcal{D}=\frac{1}{2}T_A T_B C_{AB}^K\partial_{K}$,
   One can easily obtain all our corellators as derivatives of $\mathcal{F}$.
   We now prove this equation in the generic case.

  Let us introduce an operator-valued differential form of indefinite degree on $R$
  \be
  U=e^{-tH}+G_+dte^{-tH}
  \ee
  We can check that
  \be
  dU+\{Q,U\}=0
  \ee
  One can construct the differential form on $R^{k-1}$ which takes values in $Hom (V_0,V_0)$
  and includes $k+1$ $\Phi$'s
  \be
  \omega^{(k-1)}=\Pi_0 \Phi U_1...   \Phi U_k  \Phi \Pi_0  \\
  \ee
  Our form is $d$-closed because $Q\Pi_0=\Pi_0Q=0$ and $\Phi$ anticommutes with $Q$
\be
d\omega^{(k-1)}=-\Pi_0 \Phi \{U_1,Q\}...   \Phi U_k  \Phi \Pi_0-
..-\Pi_0 \Phi U_1...   \Phi \{U_k,Q\}  \Phi \Pi_0=0
\ee
   One can consider the integral of the degree $k-1$ component of our form
 over the boundary of some surface and rewrite it as
an integral of $d\omega^{(k-1)}$ over this surface by the Stocks theorem.
  The surface of integration is the boundary of the $k$-dimensional cube.
The result of integration is
  as follows
  \be
\int ^{\infty}_0..\int ^{\infty}_{0} \Pi_0 \Phi (\Pi_0 - 1) \Phi
G_+ e^{-t_1H}... G_+e^{-t_kH}\Phi \Pi_0 dt_1..dt_{k-1} +
(perm.\   \ of\   \  \Pi_0-1  ) = 0 \ee
 The terms that include  $\Pi_0$ come from the commutator $\{ \mathcal{F}, \mathcal{F}\}$, while
the other terms come from $T_A T_B C_{AB}^K\partial_{K}\mathcal{F} $.
 For our superalgebra of operators $\Phi_A$, one can write (super)Jacobi identities in the following form
\be
C_{AB}^{K}C_{KD}^{E}=0 \ \ , \ \ \mathcal{D}^2=0
\ee
Our anticommutativity equation  can be rewritten as the zero curvature
equation \be\label{u}
\{\mathcal{D}-\mathcal{F},\mathcal{D}-\mathcal{F}\}=0 \ee

\noindent {\bf 4.} Now we can illustrate how this equation (\ref{u}) works
in the simplest case. Define in (\ref{f1}) $\mathcal{F}^{(1)}=\sum F_AT_A $. Then
\be \{F_A,F_B\}=C_{AB}^{K}F_{K} \ee
For odd operators
$\Phi_A$'s, this equation is quite nontrivial. One can see this by
considering the matrix example. In the simplest case, our space of states
$V$ is the four-dimensional vector space. The operators $Q$ and
$G_+$ are $4\times 4$ matrices \be Q=\left(
\begin{array}{cccc}
0&0&0&0 \\
1&0&0&0 \\
0&0&0&0 \\
0&0&0&0 \\
\end{array}
\right)
               \     \
G_+= \left(
\begin{array}{cccc}
0&1&0&0 \\
0&0&0&0 \\
0&0&0&0 \\
0&0&0&0 \\
\end{array}
\right)                H=\{Q,G_+\}=\left(
\begin{array}{cccc}
1&0&0&0 \\
0&1&0&0 \\
0&0&0&0 \\
0&0&0&0 \\
\end{array}
\right)

\ee $V_0$ is the subspace of $V$ with first two zero coefficients.
The operators $\Phi_A$'s which satisfy commutation relation $(15)$
are as follows \be \Phi_A=\left(
\begin{array}{cc}
 \begin{array}{cc}
  a&0  \\
  **&-a \\
 \end{array} &
\begin{array}{cc}
  0&0  \\
  **&* \\
 \end{array} \\
 \begin{array}{cc}
  *&0  \\
  **&0 \\
 \end{array} &
 F_A\\
\end{array}
\right)
\ee
where stars are some numbers.
 When considering commutation relations on $\Phi_A$, we are interested only
in the right bottom block. It is easy to show that in this block there is anticommutator of the two right
bottom blocks of $\Phi_A$ and $\Phi_B$.
It is what we get commuting matrix elements $F_A$

\be
\{\Phi_A ,\Phi_B\}=\left(
\begin{array}{cc}
 \begin{array}{cc}
  *&*  \\
  **&* \\
 \end{array} &
\begin{array}{cc}
  *&* \\
  **&* \\
 \end{array} \\
 \begin{array}{cc}
  *&*  \\
  **&* \\
 \end{array} &
 \{F_A,F_B\}\\
\end{array}
\right) \ee
Thus, we observe that this is only these special commutation relations of matrices $\Phi_A$
that provide us with the similar algebra for $F_A$ (i.e. for the right bottom
blocks).

Now we can explicitly show this for general $V$. Consider \be\label{cc}
\{F_A,F_B\}=\Pi_0\Phi_A\Pi_0\Phi_B\Pi_0+\Pi_0\Phi_B\Pi_0\Phi_A\Pi_0
\ee Using formula $(15)$ for the projector, one arrives at the
following formula \be\label{ccc} \left.\Pi_0=\lim_{t \rightarrow
\infty}e^{-tH}=\lim_{t \rightarrow \infty}\sum
(-t)^k\frac{H^k}{k!}=
1+\left(\sum QG_+ (-t)^k\frac{H^{k-1}}{k!}+\sum (-t)^kG_+\frac{H^{k-1}}{k!}Q\right)\right|_{t \rightarrow \infty} \\
\ee
An important property of this representation is
 \be
\Pi_0\Phi_AQG_+(-t)^k\frac{H^{k-1}}{k!}\Phi_B\Pi_0=0\\
\Pi_0\Phi_AG_+(-t)^k\frac{H^{k-1}}{k!}Q\Phi_B\Pi_0=0\\
\ee
Hence, the only term surviving in the expression for the projector in (\ref{cc})  is the $c$-number
term in (\ref{ccc}) and \be
\{F_A,F_B\}=\Pi_0\Phi_A\Phi_B\Pi_0+\Pi_0\Phi_B\Phi_A\Pi_0=C_{AB}^{K}F_{K}
\ee

\noindent {\bf 5.} Our anticommutativity equation contains a
commutativity equation introduced in [5,6,7,8]\footnote{This equation is contained as the $t$-part
in the $t-t^{*}$ equations of [5]. }  for
special operators.
 The commutativity equation has the form
 \be
{[B_\mu,B_\nu]}=0 \    \ , \
\ B_\mu=\frac{\partial B}{\partial\tau_\mu}
\ee
Indeed, consider a new odd operator $G_-$ that satisfies
\be\label{relg}
G_-^2=0 \ \ , \ \ \{G_-,G_+\}=0 \   \ ,
\  \ \{G_-,Q\}=0 \ \ , \ \ G_-V_0=0
\ee
Consider even operators $\Phi_\mu$ which satisfy relations (\ref{relf}) with $C=0$ and
 \be
{[{[\Phi_\mu,G_-]},\Phi_\nu]}=0
 \ee
Thus properties allow one to
add the new odd operators $\Phi'_{\mu}={[\Phi_\mu,G_-]}$ to our
algebra because they satisfy (\ref{relf}) and write $\mathcal{F}$
in these terms. Let us
introduce the two kinds of coupling constants $T_A$: even $\tau_\mu$ for odd
operators ${[\Phi_\mu,G_-]}$
and odd $\theta_\mu$ for even $\Phi_\mu$.

 The main object in the commutativity equations in
terms of $\mathcal{F}$ is
\be
 B_\mu=\frac{\partial\mathcal{F}}{\partial\theta_\mu}|_{\theta=0}
\ee
 In fact, one can show that  (\ref{relg}) leads to
\be\label{www}
\mathcal{F}|_{\theta=0}=0 \  \ , \  \ \hbox{hence} \\
\frac{\partial}{\partial\theta_\mu}\frac{\partial}{\partial\theta_\nu}
\{\mathcal{F},\mathcal{F}\}|_{\theta=0}=
{[\partial_\mu\mathcal{F}|_{\theta=0},
 \partial_\mu\mathcal{F}|_{\theta=0}]}=0
\ee
The equation (\ref{www}) would  be the  commutativity equation if one demonstraates that
$B_\mu=\frac{\partial B}{\partial \tau_\mu}$. It follows from the
properties of $G_-$
 \be
 B_\mu=\frac{\partial B}{\partial\tau_\mu} \Longrightarrow
 \frac{\partial B_\nu}{\partial\tau_\mu}=
\frac{\partial B_\mu}{\partial\tau_\nu} \sim \\
..\Phi_\mu G_+e^{-tH} {[\Phi_\nu,G_-]}...= ..{[\Phi_\mu,G_-]} G_+e^{-tH}\Phi_\nu...
\ee
\noindent {\bf 6.}
 Consider the deformation of
solutions to the anticommutativity equation by the variation of operator
$G_+$ in terms of $\mathcal{K}=\frac{G_+}{H}=\int_0^\infty
G_+e^{-tH}dt$  so that
 \be
 \{\mathcal{K},Q\}=1-\Pi_0
 \ee
The variation of $\mathcal{K }$ satisfies
\be
\{\delta\mathcal{K},Q\}=0
\Longrightarrow  \delta\mathcal{K}= {[Q,Z]}
\ee
The variation $\delta\mathcal{K}$ is exact, since $Q$ does not have cohomologies in $V_1$.
 We can write the variation of the  solution to the anticommutativity equation
 and after some algebra we get
 \be\label{vari}
\delta_{\mathcal{K}}\mathcal{F}=-\{\mathcal{D}-\mathcal{F},\mathcal{F}_Z\}_s
\ee
In (\ref{vari}) $\mathcal{F}_Z$ is obtained from the expression\footnote {
 Schematically, $\mathcal{F}=\sum \Pi_0\Phi\mathcal{K}\Phi\mathcal{K}...\Phi\mathcal{K}\Phi\Pi_0$.}
for $\mathcal{F}$ in terms of $\mathcal{K}$
by replacing  insertions of $\mathcal{K}$ by  $Z$. The
variation of  $\mathcal{F}$ in the form (\ref{vari}) is a counterpart  of
the gauge transformations in the zero curvature equation (\ref{u})
because they retain this equation
\be
\{\mathcal{D}-\mathcal{F},\mathcal{D}-\mathcal{F}\}=0
\Longrightarrow \{\mathcal{D}-\mathcal{F},\delta\mathcal{F}\}=0 \\
\{\mathcal{D}-\mathcal{F},\delta_{\mathcal{K}}\mathcal{F}\}=
\{\mathcal{D}-\mathcal{F},\{\mathcal{D}-\mathcal{F},\mathcal{F}_Z\}_s\}=0
\ee

The special property of commutativity equation is that $B_\mu$ is
invariant under the variations of $\mathcal{K}$, it follows from
(\ref{vari}) and (\ref{relg}). Actually $\mathcal{F}|_{\theta=0}=0$
and $\mathcal{F}_Z|_{\theta=0}=0$, hence the variation term linear in
$\theta$ is
\be
 \delta_{\mathcal{K}}B_\mu= \delta_{\mathcal{K}}
\frac{\partial \mathcal{F}}{\partial\theta_\mu}|_{\theta=0}=0
\ee
In the case of anticommutativity equation one generally has only two
terms invariant  under the variations of $\mathcal{K}$:
$\mathcal{F}^{(1)}$ and $\mathcal{F}^{(2)}$.
\\

{\bf Acknowledgements:}
The author is grateful to A. Morozov, A. Mironov and especially to A. Losev
 for fruitful discussions. The author also acknowledges the kind hospitality in
Kaiserslautern University  during the preparation of
this work. This work is partly supported by grant RFBR 01-02-17682-a and by the
Volkswagen-Stiftung.
\\

{\bf References:}

[1]E.Witten,"On the structure of the topological phase of
 two-dimensional gravity", \\ Nucl.Phys.B340:281-332, 1990

[2]E.Witten,
"Quantum Field Theory and the Jones Polynomial",  \\ Commun.Math.Phys.121:351,1989

[3]E.Witten,"Topological quantum field
theories",Commu.Math.Phys.117:353,1988

[4] E.Witten,"Topological sigma models",Commun.Math.Phys.118:411,1988

[5] S. Cecotti , C. Vafa,
"Topological antitopological fusion",
Nucl.Phys.B367:359-461,1991

[6]A.Losev,"Commutativity equations, operator-valued cohomologies
of the "sausage"  compactification of $(C^*)^N/C^*$ and SQM"
Preprint ITEP-TH-84/98, LPTHE-61/98.

[7]A.Losev and  Yu.Manin, "New moduli space of pointed curves
 and family of flat connections", AG/0001003

[8]A.Losev and I. Polyubin,
 "On compatibility of tensor products on solutions
 to commutativity and WDVV equations",
 Pisma Zh.Eksp.Teor.Fiz.73:59-63,2001

 \end{document}